\documentclass[reprint, prb, superscriptaddress]{revtex4-1}

\usepackage[pdftex]{graphicx}

\begin{document}

\title{Hydrated lithium intercalation into the Kitaev spin liquid candidate material $\alpha-$RuCl$_3$}%

\affiliation{Department of Physics, Tohoku University, 6-3 Aramaki-Aoba, Aoba-ku, Sendai, Miyagi 980-8578, JAPAN}

\author{Yoshinori Imai}
\email{imai@tohoku.ac.jp}
\author{Katsuya Konno}
\author{Yoshinao Hasegawa}
\author{Takuya Aoyama}
\author{Kenya Ohgushi}
\affiliation{Department of Physics, Tohoku University, 6-3 Aramaki-Aoba, Aoba-ku, Sendai, Miyagi 980-8578, JAPAN}

\newcommand{\ru}{RuCl$_3$}
\newcommand{\mizu}{H$_2$O}
\newcommand{\liru}{Li$_x$RuCl$_3$}
\newcommand{\liruh}{Li$_x$RuCl$_3 \cdot y$H$_2$O}
\newcommand{\tn}{$T_\mathrm{N}$}
\newcommand{\ts}{$T^*$}
\newcommand{\eg}{$E_g$}

\begin{abstract}
We study on transport and magnetic properties of hydrated and lithium-intercalated $\alpha$-\ru, \liruh, for investigating the effect on mobile-carrier doping into candidate materials for a realization of a Kitaev model.
From thermogravitometoric and one-dimensional electron map analyses, we find two crystal structures of this system, that is, mono-layer hydrated \liruh~$(x\approx0.56, y\approx1.3)$ and bi-layer hydrated \liruh~$(x\approx0.56, y\approx3.9)$.
The temperature dependence of the electrical resistivity shows a temperature hysteresis at 200-270 K, which is considered to relate with a formation of a charge order.
The antiferromagnetic order at 7-13 K in pristine $\alpha$-\ru~ is successfully suppressed down to 2 K in bi-layer hydrated \liruh, which is sensitive to not only an electronic state of Ru but also an interlayer distance between Ru-Cl planes.
\end{abstract}

\maketitle
\begin{table*}[bht]
\caption{Specification of samples used in this study.
``MLH-'' or ``BLH-'' in composition represents that the sample is a mono-layer hydrate or a bi-layer hydrate, respectively.  
Details of ``method 1'' and ``method 2'' in the post process are described in the text.
The lattice parameters $(a,~b,~c,~\beta)$ are deduced on an assumption of monoclinic space group $C2/m$.
The $c^*$ represents the interlayer distance of the Ru-Cl plane.
In the column of antiferromagnetic transition temperature~(\tn), ``$<2$ K'' represents that an antiferromagnetic transition does not appear above 2 K in the magnetic susceptibility measurements.
}
\begin{tabular}{ccccccccccc}
\hline
& composition & form & solvent & post process & $c^*$ (\AA) & $a$ (\AA) & $b$ (\AA) & $c$ (\AA) & $\beta$ (deg.) & $T_\mathrm{N}$ (K)\\ \hline
sample A & \ru & polycrystal & - & - & 5.72 & 5.98 & 10.36 & 6.04 & 108.9 & 13.2 K\\
sample B & BLH-\liruh & polycrystal & ethanol & - & 11.12& 6.04 & 10.43 & 11.16 & 90.1 & $< 2$ K\\
sample C & BLH-\liruh & polycrystal & 2-propanol & - & 10.95 & 6.04 & 10.48 & 11.02 & 90.1 & $< 2$ K\\
sample D & MLH-\liruh & polycrystal & ethanol & method1 & 8.17 & 6.03 & 10.35 & 8.25 & 98.3 & 3.6 K\\
sample E & BLH-\liruh & polycrystal & ethanol & method2 & 11.22 & 6.03 & 10.42 & 11.27 & 90.0 & $< 2$ K\\
sample F & \ru & single crystal & - & - & 5.73 & - & - & - & - & $7.5$ K, $13.2$ K\\
sample G & MLH-\liruh & single crystal & ethanol & method1 & 8.23 & - & - & - & - & $3.6$ K\\
sample H & BLH-\liruh & single crystal & ethanol & method2 & 10.98 & - & - & - & - & $< 2$ K\\ \hline
\end{tabular}
\label{tab:spec}
\end{table*}
\begin{figure}[b]
\begin{center}
\includegraphics[width=0.99\linewidth]{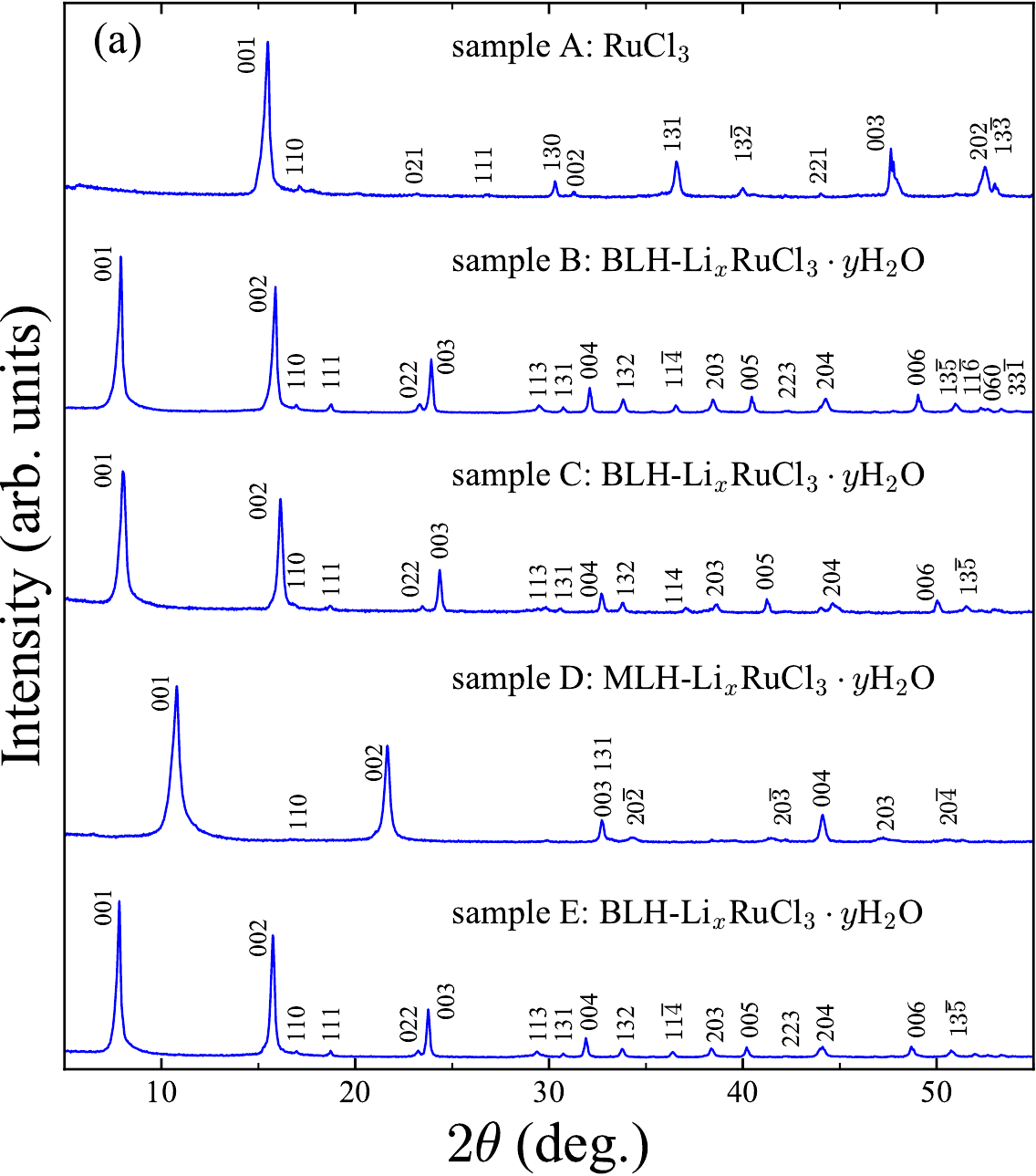}\\
\vspace{1mm}
\includegraphics[width=0.99\linewidth]{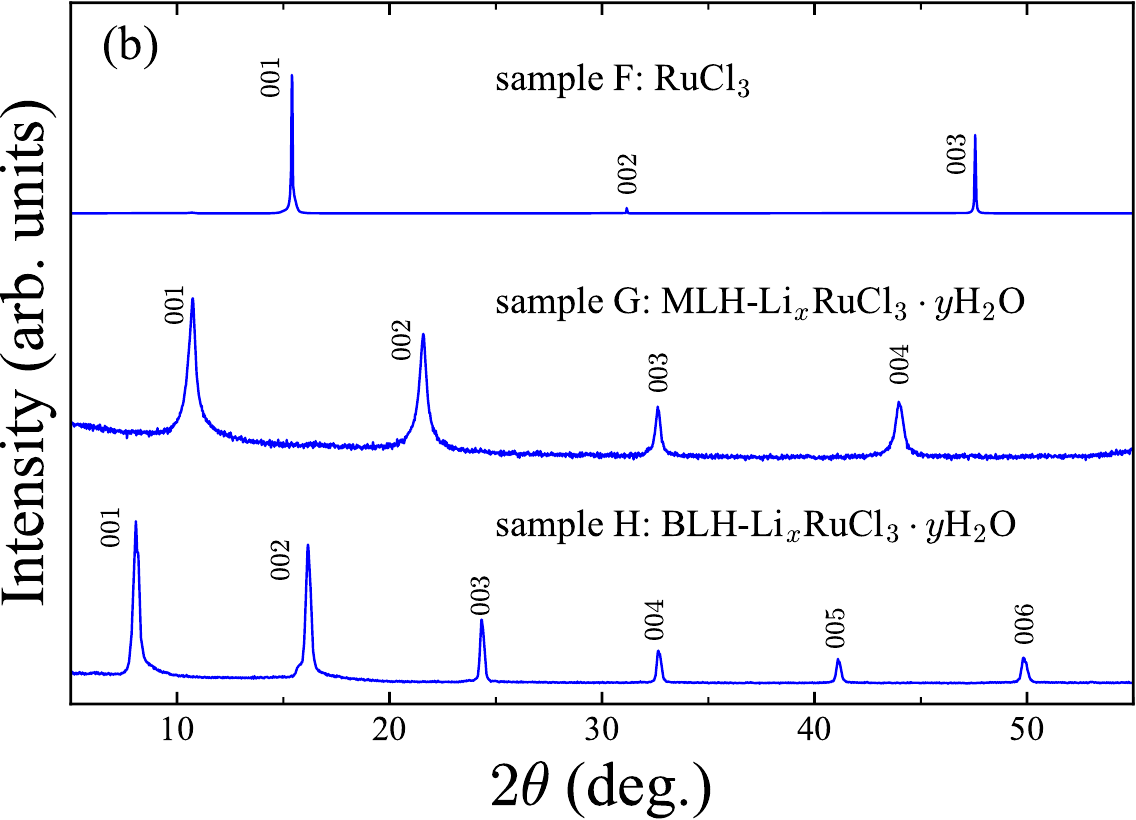}
\caption{
X-ray diffraction patterns for (a) polycrystals of samples A-E and (b) single crystals of samples F-H.
The vertical axis is in a logarithmic scale.
The Miller indexes on the basis of the monoclinic $C2/m$ are also shown.
}
\label{fig:xrd}
\end{center}
\vspace{-5mm}
\end{figure}
\begin{figure}[b]
\begin{center}
\includegraphics[width=1\linewidth]{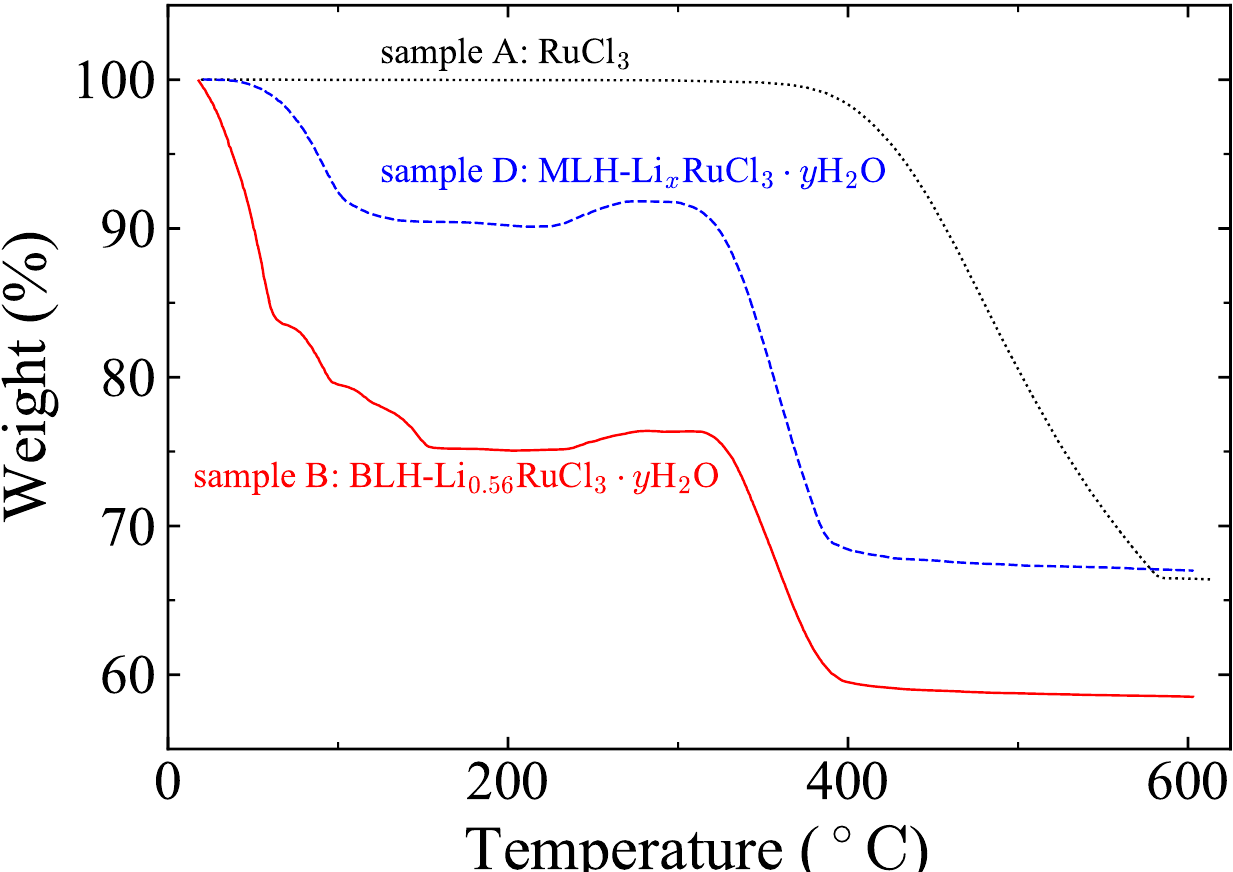}
\caption{
Thermogravimetric (TG) analysis for samples A, B, and D in the heating process up to $600^\circ \mathrm{C}$ at $1^\circ \mathrm{C} /\mathrm{min.}$.
Temperature dependence of weight loss is shown.}
\label{fig:tg}
\end{center}
\vspace{-5mm}
\end{figure}
\begin{figure}[htb]
\begin{center}
\includegraphics[width=1\linewidth]{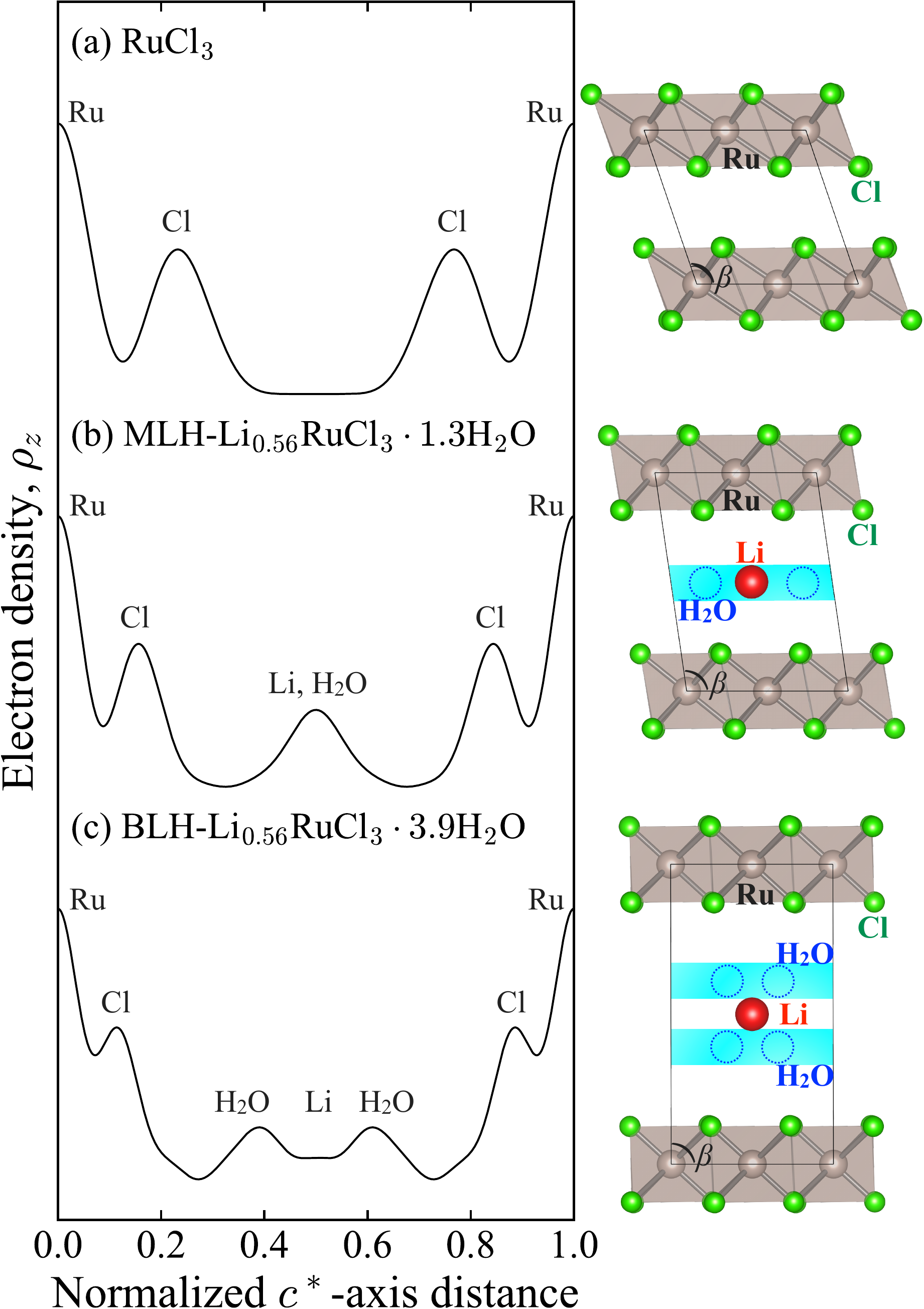}
\caption{
One-dimensional electron density map, $\rho_z$, and the structural model of $ac$-plane for single crystals of (a) \ru~(sample F), (b) MLH- and (c) BLH- \liruh~(samples G, H).
}
\label{fig:ed}
\end{center}
\vspace{-5mm}
\end{figure}
\begin{figure}[htb]
\begin{center}
\includegraphics[width=1\linewidth]{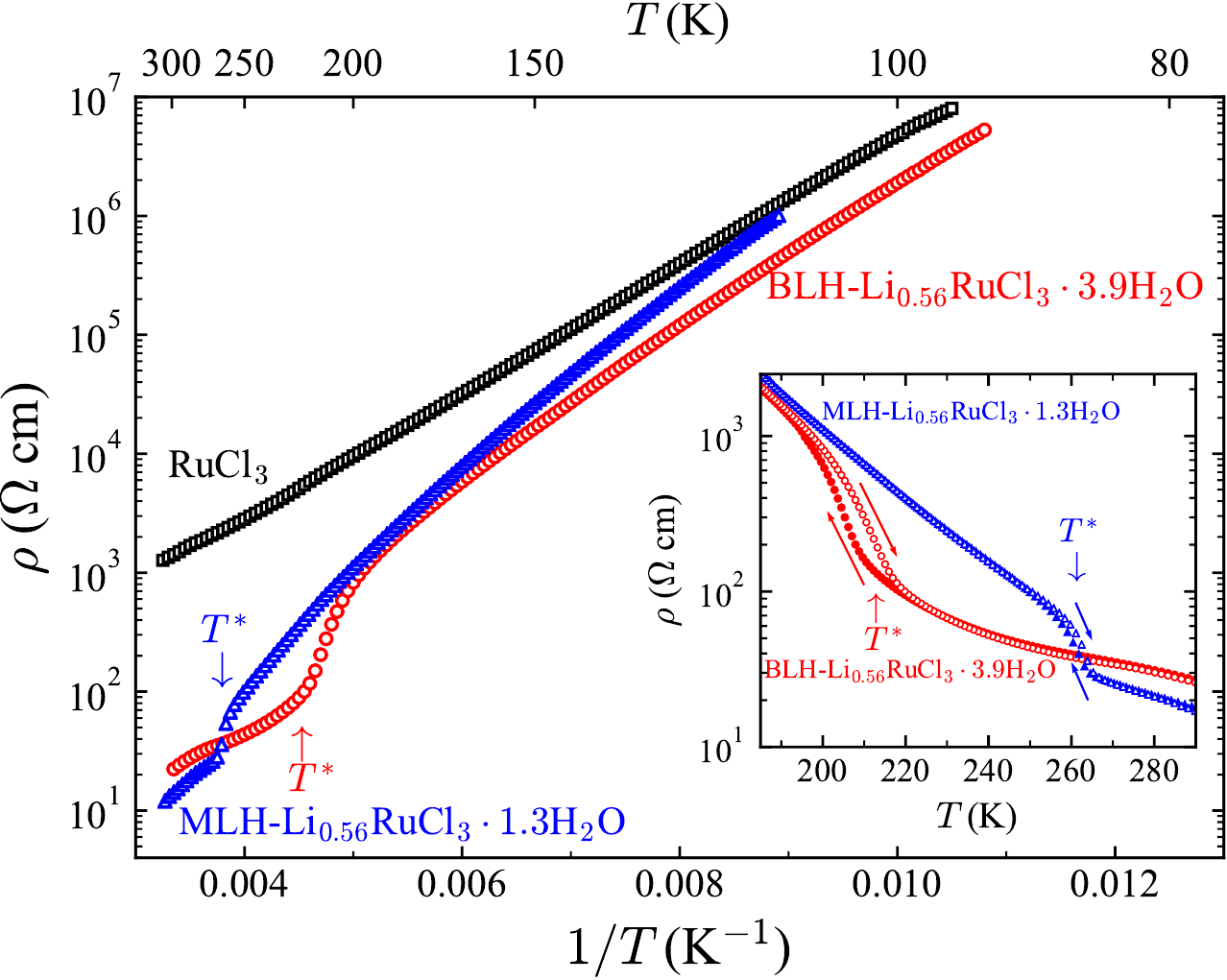}
\caption{
Arrhenius plot of the in-plane resistivity, $\rho$, for single crystals of \ru, MLH- and BLH- \liruh~(samples F-H).
The inset shows the resistivity in cooling and warming cycles, which are represented by closed and open symbols, respectively, in the vicinity of the phase transition temperature, \ts.
}
\label{fig:rho}
\end{center}
\vspace{-5mm}
\end{figure}
\begin{figure}[h]
\begin{center}
\includegraphics[width=0.78\linewidth]{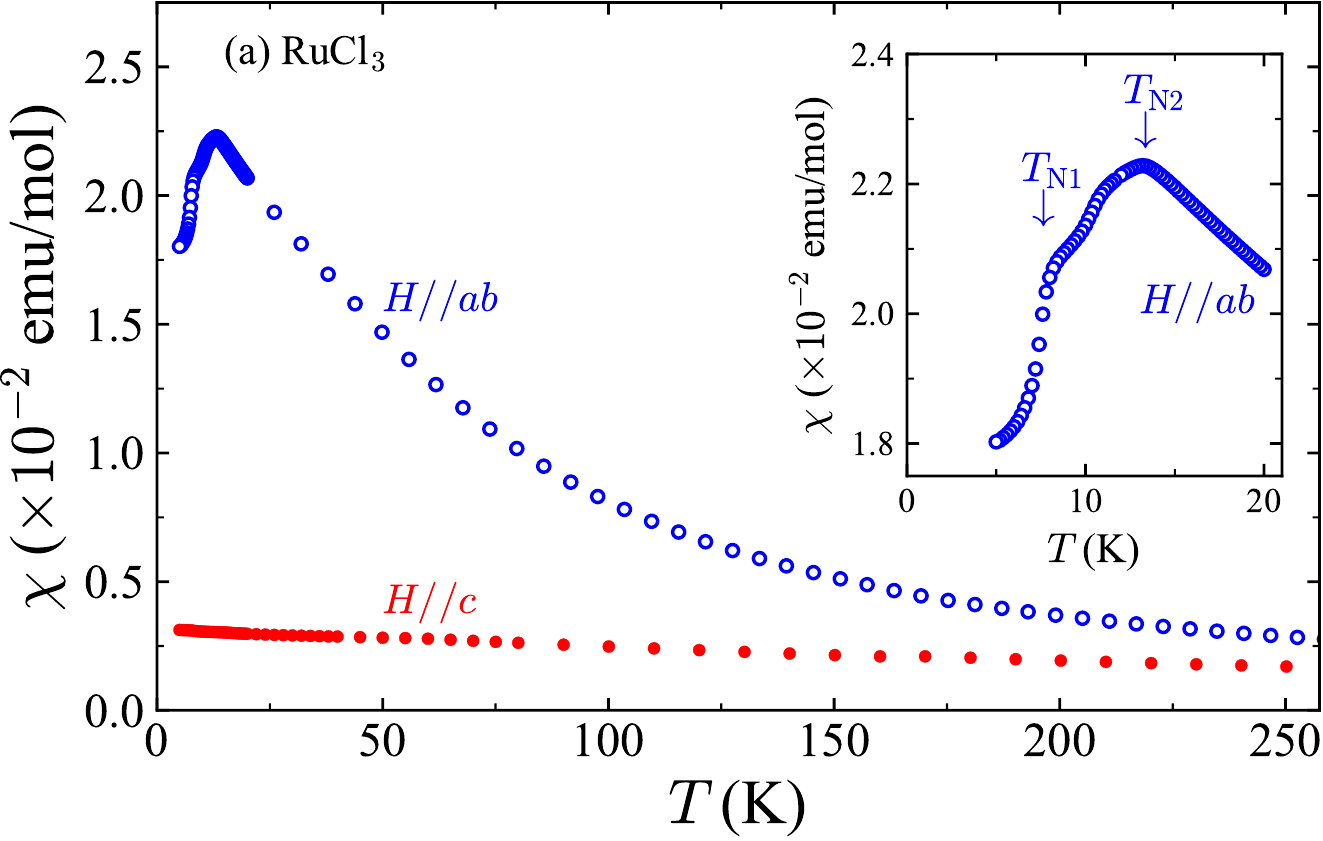}\\
\vspace{1mm}
\includegraphics[width=0.78\linewidth]{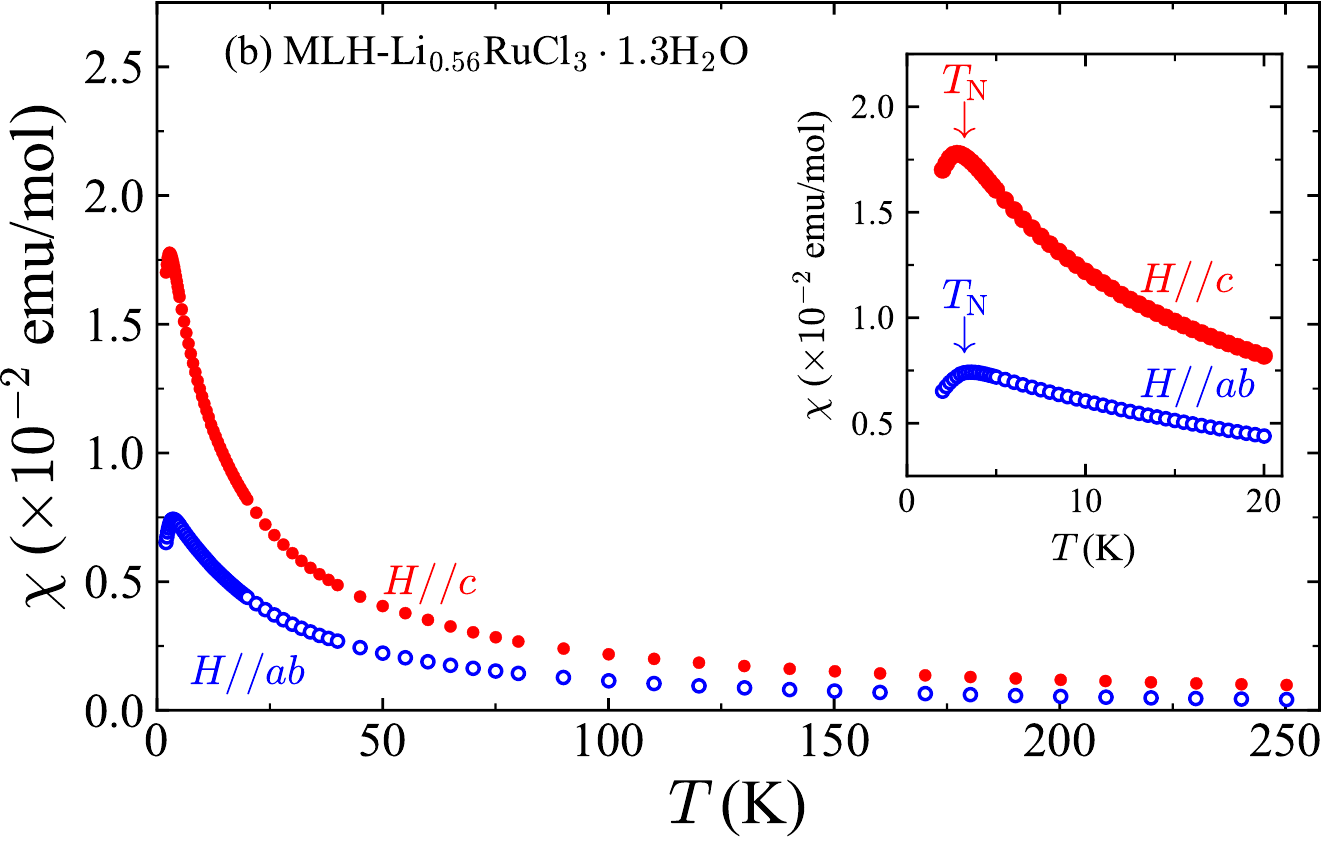}\\
\vspace{1mm}
\includegraphics[width=0.78\linewidth]{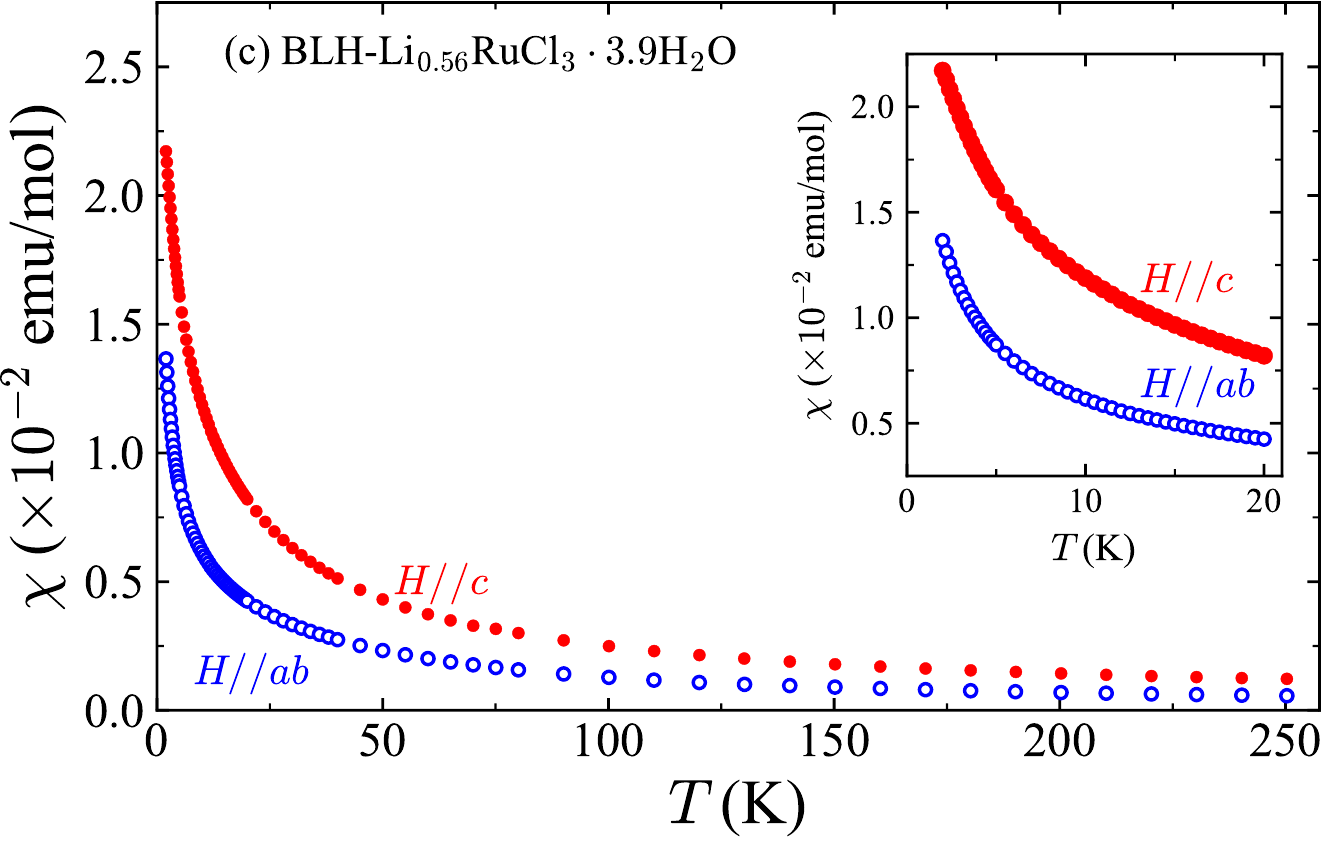}\\
\vspace{1mm}
\includegraphics[width=0.78\linewidth]{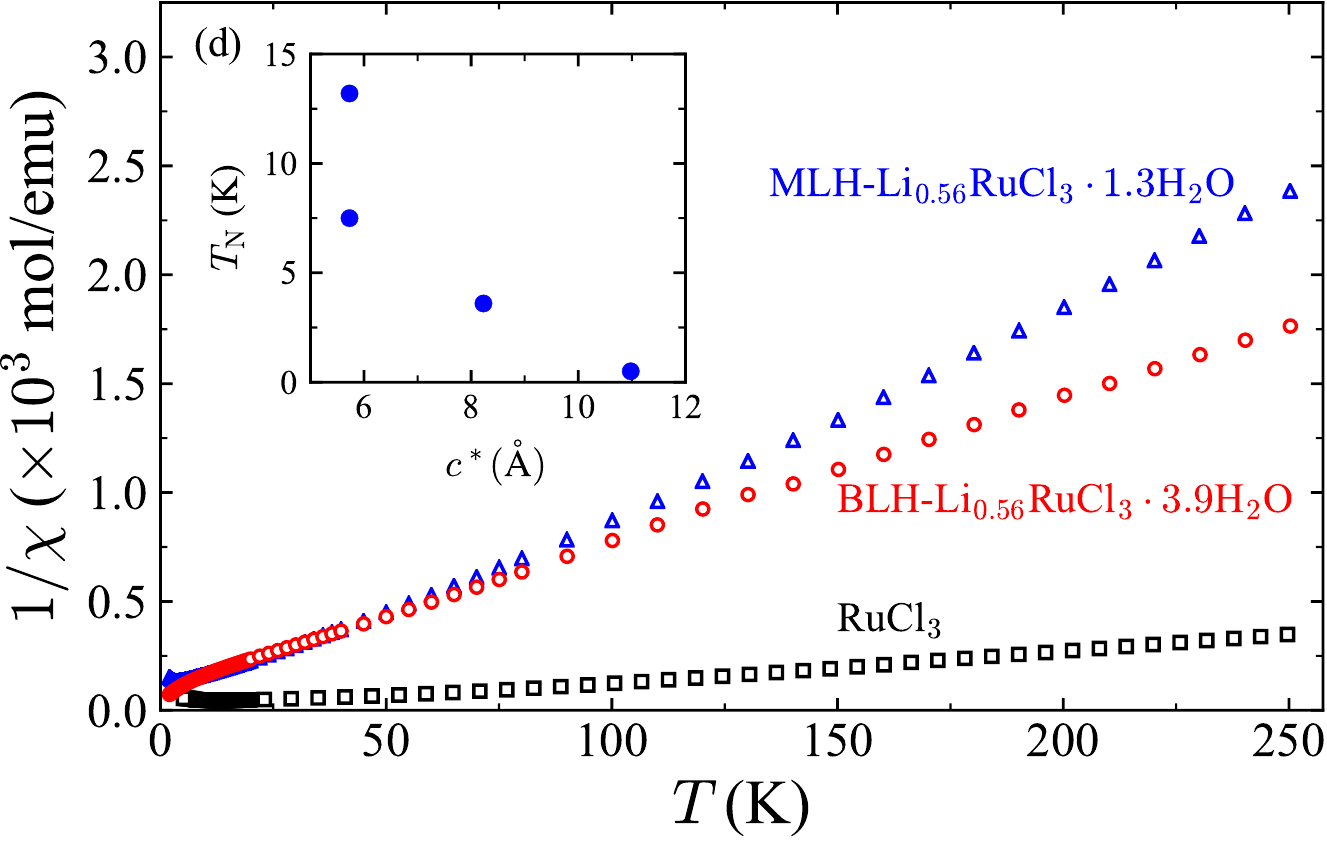}
\caption{
Temperature dependences of magnetic susceptibility, $\chi$, of single crystals of (a) \ru, (b) MLH-, and (c) BLH- \liruh~at a magnetic field of $\mu_0 H = 1$ T parallel to the $ab$-plane (open symbols) and the $c$-axis (closed symbols).
Insets are enlarged plots at low temperatures.
(d) Curie-Weiss plot of $\chi$ under the magnetic fields parallel to $ab$-plane.
The inset of (d) is a dependence of an antiferromagnetic transition temperature, \tn, on an interlayer distance, $c^*$.
}
\label{fig:chi}
\end{center}
\vspace{-8mm}
\end{figure}
\section{Introduction}
Triggered by a proposal of new quantum model called the Kitaev model \cite{Kitaev}, the tremendous numbers of studies have been performed on a quantum spin liquid, especially a Kitaev quantum spin liquid \cite{JPCM.29.493002,ARCMP.9.17,NRP.1.264,ARCMP.10.451}.
The Kitaev model is a very simple model where $S=1/2$ spins are placed on a honeycomb lattice and are coupled with a nearest-neighbor bond-dependent interactions.
The most remarkable feature of the Kitaev model is that this is an exactly solvable model, which shows that the ground state is the Kitaev quantum spin liquid and that Majorana fermions emerge as excitations~\cite{PRL.112.207203,PRL.113.187201,PRB.92.115122}.
Since bond-dependent interactions naturally exist in materials with strong spin-orbit couplings  \cite{PTPS.160.155,PRL.102.017205}, some compounds with an unfilled $4d/5d$ orbitals have been attracting intensively \cite{NRP.1.264}.
Especially,  $\alpha$-\ru~is the most probable candidate material for the Kitaev quantum spin liquid, since $J_{eff}=1/2$ spins are coupled with each other through the Kitaev-type ferromagnetic interactions~\cite{PRB.90.041112}. 

The space group of \ru~is $C2/m$ \cite{PRB.92.235119}, and honeycomb lattices of octahedrally coordinated Ru$^{3+}$ ions are stacked via a van der Waals interaction.
The Ru$^{3+}$ ions have low-spin configuration of $(t_{2g})^5$, bearing effective $J_{eff}=1/2$ spins.
Contrary to expectations from the Kitaev model, \ru~shows an antiferromagnetic~(AF) transition around an AF transition temperature, $T_\mathrm{N}=7-13$ K, which is considered to be due to non-Kitaev interactions, such as direct exchange interactions and next-nearest neighbor superexchange interactions.
Recent investigations assigned the phase with $T_\mathrm{N} \sim 7$ K to an $ABC$ stacking order and that with $T_\mathrm{N} \sim 13$ K to an $AB$ stacking fault \cite{PRB.92.235119,PRB.93.134423}.
However, upon the application of in-plane magnetic fields, an antiferromagnetic order is fully suppressed down to the lowest temperature, and the half-integer quantization is observed in the thermal Hall conductance measurement, which provides a direct evidence for capturing Majorana fermions \cite{Nat.559.227}.

The study on the substitution effect for \ru~is intriguing to reveal the role of impurities for a realization of Kitaev model, and earlier studies on (Ru$_{1-x}$Ir$_x$)Cl$_3$ clarified that the spin liquid like state appears in the wide range of an electronic phase diagram \cite{PRL.119.237203, PRB.98.014407}.
Introduction of not localized impurities but rather mobile charge carriers into \ru~is a more challenging issue \cite{PRL.119.237203, PRB.98.014407,PRM.1.052001,JACS.122.6629, nanolett.16.3578}, because some theoretical studies predict an emergence of novel superconductivity in the carrier-doped Kitaev material  \cite{PRB.85.140510,PRB.86.085145,PRL.108.227207,PRL.110.066403,PRB.87.064508,PRB.90.024404,PRB.91.220501,PRB.97.014504}.
In an electron-doped material K$_{0.5}$RuCl$_3$~(a formal valence is Ru$^{2.5+}$ with the $(4d)^{5.5}$ electron configuration), which are prepared by K-coating on a \ru~single crystal cleaved in a vacuum chamber, a charge order of $(4d)^5$ and $(4d)^6$ states is proposed at low temperatures \cite{PRM.1.052001}.
Li-intercalated material \liru~prepared by using LiBH$_4$ reveals that the antiferromagnetic order is suppressed below 2 K, and that the electrical resistivity still remains an insulating behavior \cite{JACS.122.6629, nanolett.16.3578}.
In the study, the Li content, $x$, is as low as $x=0.2$, which is smaller than the honeycomb-lattice percolation threshold, $x_p=0.303$.
Therefore, the studies over a wide carrier concentration range are highly expected for understanding the doping effect on Kitaev materials.

Here, we report on the successful preparation of hydrated and Li-intercalated \ru, in which electron carriers are doped into \ru~by using a soft-chemical method, and the investigation of their electronic properties.
In \liruh, there are two kinds of crystal structures, $i.e.$, mono-layer hydrate~(MLH) and bi-layer hydrate~(BLH).
The AF state is completely suppressed down to 2 K in BLH-\liruh.
It turned out that \tn~depends on an electronic state of Ru as well as the distance between Ru-Cl layers.

\section{Experimental}
We prepare 8 samples of $\alpha$-\ru,~and hydrated and Li-intercalated \ru, whose detailed specifications are summarized in Table \ref{tab:spec}.
Commercially available $\alpha$-\ru~polycrystalline powders (3N, Mitsuwa Chemicals) were used as a pristine sample in this study, which is represented as sample A in Table \ref{tab:spec}.
Hydrated Li-intercalated samples are prepared as follows.
\ru~powders of 0.3 g were soaked in 1.5 mol$/l$ LiI solution of ethanol~(2-propanol), which contain a few percents of water, at their boiling point for 2 hours.
This reaction can be described by the following chemical reaction formula,
$
\mathrm{RuCl}_3 + x\mathrm{LiI} + y\mathrm{H}_2\mathrm{O} \to \mathrm{Li}_x\mathrm{RuCl}_3 \cdot y \mathrm{H}_2\mathrm{O} + \frac{x}{2}\mathrm{I}_2.
$
Then, the samples are washed by the same liquid as a solvent, and dried at room temperature, which are ``sample B''~(``sample C'').
For clarifying whether \mizu~is actually intercalated into samples, we tried two kinds of post process for sample B.
At first, powders of sample B are kept with silica gel in a sealed vessel for 1 day; this process is called as ``method 1''.
The obtained sample is named as ``sample D'' in Table \ref{tab:spec}.
Next, we store sample D with a wet cotton in a sealed vessel for 1 day; this process is called ``method 2''. 
The product is named as ``sample E'' in Table \ref{tab:spec}.
\ru~single crystals~(Sample F) were prepared by the chemical vapor transport method as described elsewhere \cite{hase17,PRB.95.245104}.
Some pieces of \ru~single crystals~(typical size:~$2\times2\times0.1$~mm$^3$) were soaked in 1.5 mol$/l$ LiI solution of ethanol at room temperature for 24 hours, and then washed by ethanol before drying at room temperature.
After then, intercalated crystals were kept in a sealed vessel with silica gel or a wet cotton for 1 day, which are ``sample G'' and ``sample H'' in Table \ref{tab:spec}, respectively.

All the products were characterized by the powder X-ray diffraction (XRD) using the Cu $K\alpha$ radiation at room temperature.  
The chemical composition was determined by the inductively coupled plasma atomic emission spectroscopy (ICP-AES) and thermogravimetry~(TG) analysis.
The electrical resistivity, $\rho$, was measured by the four-terminal method over the temperature range of 77 to 300 K.
The current direction is along the $ab$-plane.
Magnetic susceptibility measurements were performed using a superconducting quantum interference device (SQUID) magnetometer.

\section{Results}
Figure \ref{fig:xrd}(a) shows XRD patterns for polycrystalline samples A-E.
All peaks can be indexed on the basis of a monoclinic space group (No. 12, $C2/m$) \cite{PRB.92.235119}, and calculated lattice parameters are summarized in Table \ref{tab:spec}.
The peak positions of $00l$ peaks for intercalated samples shift toward smaller $2\theta$ values than those for a pristine sample~(sample A), indicating the successful intercalation.
It is highly unlikely that this considerable increase in the interlayer distance, $c^*$, between Ru-Cl layers results from the intercalation of only Li ions, because in the other Li-intercalated materials such as Li$_{x}$TaS$_2$ and Li$_{x}$NbS$_2$, the increase of interlayer distance is known to be as small as 1~\AA~\cite{IC.16.2950,MRB.14.797}.
Thus, it is quite reasonable that some kinds of molecules are co-intercalated with Li ions.
Here, it should be noted that the peak positions of samples B and C, which are respectively synthesized in the ethanol and 2-propanol as  solvents, are almost the same, indicating that the intercalated molecule is the same one in samples B and C.
The most probable candidate of the intercalated molecule in both sample B and sample C is \mizu, which is included in both of ethanol and 2-propanol.
For clarifying whether \mizu~molecule is actually co-intercalated with Li ions in samples B and C, we tried two kinds of post process described above for sample B.
In XRD patterns of sample D, which had been kept with silica gel through the post process of method 1, the positions of $00l$ peaks shift towards larger $2\theta$ values; the $c^*$ value decreases by $\sim 3$ \AA~from $c^*$ of sample B.
Interestingly, when sample D is kept under high humidity for 1 day through the post process of method 2, the positions of $00l$ shift towards smaller $2\theta$ values again, and the resultant XRD pattern of sample E is the same as that of sample B.
This shows that the intercalated molecules exist not only in the solvent but also in air, which indicates that the co-intercalated molecule is \mizu.
We can then conclude that there are two types of structure forms with the chemical formula \liruh~with the same $x$ value and the distinct $y$ values. 
The crystal structure with a larger (smaller) $y$ value has a longer (shorter) interlayer distance.
Comparing the lattice constants between pristine and intercalated samples, the interesting changes are observed in the parameter of $\beta$.
The angle of $\beta$ becomes close to the value of 90 degrees with increasing $c^*$, which indicates that the monoclinic distortion is relaxed by the intercalation of Li and \mizu.
Therefore, it is expected that the ideal honeycomb lattice with smaller distortion is realized in the intercalated samples compared to a pristine \ru.

In order to determine the chemical compositions $x$ and $y$ for two structural forms, we first performed the ICP analysis for sample B with a longer interlayer distance $c^*$.
This reveals that the ratio of Li and Ru is $0.56\pm0.02:1~(x=0.56)$.
Since it is likely that there is  no difference in Li concentration, $x$, between two crystal forms, we can postulate $x=0.56$ for sample D with a smaller $c^*$ value.
We then perform thermogravimetric analysis for samples A, B, and D on heating at $1^\circ$C/min in air, as shown in Fig. \ref{fig:tg}.
The weight of samples B and D decreases from room temperature to $\sim220^\circ$C, while sample A remains unchanged up to $\sim300^\circ$C.
The observed decrease in weight of samples B and D likely corresponds to a reaction of \liruh $\to$ \liru, and one can estimate \mizu~content, $y=3.9 \pm 0.1$ for sample B and $y=1.3 \pm 0.1$ for sample D.
The weight loss at temperatures higher than $\sim300^\circ$C  observed in samples A, B, and D is resulting from the decomposition and oxidization of \ru~into Ru oxides and Cl$_2$.
We consider that the intercalated single crystals (sample G, and H) take the same compositions.
Here, it should be noted that the pristine \ru~itself is stable in air.
The recent Raman spectroscopy measurements for exfoliated \ru~single crystals revealed that the Raman spectra for mono-layer single crystals of \ru~was reproducible after months of exposure to air, and that \mizu~molecule was not intercalated into \ru\cite{JPCS.2019.Henriksen}.
In addition, we confirm that soaking \ru~in ethanol does not change the lattice constant.
This is in sharp contrast to the hydrated and Li-intercalated \liruh~which changes its water content $y$ in response to changes in humidity even at room temperature. 
The moisture-sensitive behavior, which is similar to the cobalt oxyhydrate superconductor, Na$_x$CoO$_2 \cdot y$H$_2$O\cite{Nat.422.53,JSSC.177.372}, is observed only in \liruh.

For obtaining information on the location of Li and \mizu, we perform a detailed analysis of XRD patterns for single crystalline samples~(Fig. \ref{fig:xrd} (b)), where only $00l$ peaks are observed.
We obtain one-dimensional electron density (1D ED) map profiles projected along the stacking axis (defined as the $c^*$-axis).
The methodology for a calculation of 1D ED map in this study is described in details elsewhere \cite{JACS.122.6629,PRB.24.3505}.
When one considers $00l$ reflections only, the distribution of scattering density projected on the $c^*$-axis, $\rho_z$, is calculated by the Fourier summation
\begin{equation}
\rho_z = \frac{1}{c^*} \sum_l{F_{00l} \exp \left ( - i 2\pi l z \right )},
\end{equation}
in which $F_{00l}$ is the structure factor for $00l$ peaks.
For calculating $\rho_z$, the phase of $F_{00l}$ is necessary, while the absolute value of $F_{00l}$ can be estimated from the integrated intensity of $00l$ peak, $I_{00l}$, in the XRD patterns.
The phases are constrained to one of two values, that is, 0 or $\pi$, because of the centrosymmetric projection in this study, and these values are determined based on the phases of structural factors for \ru \cite{PRB.92.235119,vesta}.
This estimation is reasonable under the assumption that the contribution for the scattering from the intercalated ions or molecules is smaller than that from the \ru~component.
After the estimation of the structural model from 1D ED map, the sign of $F_{00l}$ is checked by recalculating the structural factors from the scattering of all components including intercalated atoms and molecules  \cite{vesta}.

Figure \ref{fig:ed} shows the 1D ED map of \ru~(sample F), \liruh~with $x\approx0.56,~y\approx1.3$~(sample G) and $x\approx0.56,~y\approx3.9$~(sample H).
In spite of constraint on values of phases for $F_{00l}$, 1D ED map profile of \ru~(Fig. \ref{fig:ed}(a)) is consistent with the atomic position of \ru, which indicates that the calculation method is reliable.
The 1D ED map profile for \liruh~with $x\approx0.56,~y\approx1.3$~(sample G) shows that the electron density due to guest atoms and molecules forms a single peak around the center of the gallery.
On the other hand, for \liruh~with $x\approx0.56,~y\approx3.9$~(sample H), the contributions of the intercalated atoms and molecules are observed as a small hump at the center part and two broad peaks placed 1.2 \AA~below and above the center of the gallery.
Here, we recall that there are many layered hydrates with the general formula $A_{x} \left ( MX_2 \right ) \cdot y$H$_2$O ($A=$Alkali metal, $M=$ transition metals, and $X=$O, S).
These layered hydrates generally have two kinds of crystal structures, $i.e.$, mono- and bi- layer hydrates~(MLH and BLH), where a single cation and \mizu~layer or a sequence of \mizu-cation-\mizu~layers separates the electron-doped two-dimensional $MX_2$ layers by distances of $\sim7$ \AA~or $\sim10$ \AA, respectively.
Na$_x$CoO$_2 \cdot y$H$_2$O with a triangular Co sublattice is a typical material in above series: 
BLH-Na$_x$CoO$_2 \cdot y$H$_2$O$(x \approx 0.35,~y \approx 1.3)$ shows superconductivity with a superconducting temperature of $\sim5$ K \cite{Nat.422.53}, while a superconductivity is not observed in MHL-Na$_x$CoO$_2 \cdot y$H$_2$O$(x\approx0.35,~y\approx0.7)$ \cite{JSSC.177.372}. 
We note that 1D ED maps for samples G and H in Fig.\ref{fig:ed} are similar to the electron density for MLH- and BLH- Na$_x$CoO$_2 \cdot y$H$_2$O.
In addition, the variations in $c^*$ for samples F-H shown in Table \ref{tab:spec} are similar to those in the anhydrous Na$_x$CoO$_2$, MLH- and BLH- Na$_x$CoO$_2 \cdot y$H$_2$O whose interlayer distances are 5.5, 6.9, and 9.8 \AA, respectively \cite{Nat.422.53,JSSC.177.372}.
Therefore, we conclude that sample G is MLH-Li$_{0.56}$RuCl$_3 \cdot 1.3$H$_2$O~where Ru-Cl layers are separated by a single layer of Li and \mizu, and that sample H is BLH-Li$_{0.56}$RuCl$_3 \cdot 3.9$H$_2$O~where Ru-Cl layers are separated by layers of \mizu-Li-\mizu.
Schematic pictures of crystal structures for these materials are shown in Fig. \ref{fig:ed}.
 
Figure \ref{fig:rho} shows the temperature dependence of resistivity, $\rho$, for single crystals of \ru, MLH- and BLH- \liruh~(samples F-H).
Pristine \ru~single crystal, which is a strongly spin-orbital coupled Mott insulator, shows a thermally-activated-type temperature dependence and the activation energy is $E_g \sim 0.093$ eV.
This value is lower than that reported in polycrystalline \ru \cite{JACS.122.6629}.
The resistivity for MLH- and BLH- \liruh~at room temperature is $\sim2$ orders of magnitude smaller than that of pristine \ru.
The intercalation of lithium ions makes the formal valence of Ru ions smaller than $+3$, so that the electron carriers are introduced into the material.
These electron carriers are the origin for the decrease in $\rho$ around room temperature.
The activation energy around room temperature is $E_g \sim 0.12 $ eV for MHL- and $E_g \sim 0.092 $ eV for BLH- \liruh, which are comparable with or slightly larger than that of pristine \ru.
On cooling intercalated materials, the electrical resistivity shows an anomalous hysteresis at $T^* = 260-270$ K for MHL-\liruh~and $T^* = 200-220$ K for BLH-\liruh, indicating the presence of the first-order transitions.
On further cooling below \ts, the resistivity rapidly increases and the activation energy increases up to $\sim 0.13-0.16$ eV, which is larger than \eg~of \ru.

Figure \ref{fig:chi} shows the temperature dependence of magnetic susceptibility, $\chi$, for single crystals of \ru, and MLH- and BLH- \liruh~(samples F-H) under the magnetic field of $\mu_0 H = 1$ T parallel to the $ab$-plane and the $c$-axis.
As reported previously \cite{PRB.91.144420}, in \ru~single crystals, $\chi$ for $H // ab$ is much larger than that for $H//c$, which emerges the so-called $\Gamma$ term of the spin-orbital coupling origin \cite{JPCM.29.493002}.
One can also find two magnetic transitions at $T_\mathrm{N1} \sim 7.5$ K and $T_\mathrm{N2} \sim 13.2$ K in the in-plane measurement.
Recent investigations reveal that $T_\mathrm{N1}$ is characteristics of an $ABC$ stacking ordered system, while $T_\mathrm{N2}$ is induced by the $AB$ stacking faults \cite{PRB.92.235119,PRB.93.134423}.
For $\chi$ of \ru~under $H//ab$, we perform a Curie-Weiss fit with a fitting function of $\chi = C/(T-\theta_\mathrm{CW})$ with $C=N_\mathrm{A} \mu_{eff}^2 /3 k_\mathrm{B}$ where $\theta_\mathrm{CW}$, $C$, $\mu_{eff}$, $N_\mathrm{A}$, and $k_\mathrm{B}$ are the Weiss temperature, the Weiss constant, the effective magnetic moment, the Avogadro constant, and the Boltzmann constant, respectively.
They are estimated to be $\theta_\mathrm{CW} = 25$ K and $\mu_{eff} = 2.3~\mu_\mathrm{B}/\mathrm{Ru}$, which are consistent with previous report \cite{PRB.91.144420}.
The intercalation of Li ions and \mizu~molecules results in a drastic change in magnetic properties.
The anisotropy of $\chi$ in \ru~is greatly reduced by the intercalation. 
The magnitude relationship of $\chi$ is reversed, and $\chi$ for $H//c$ is slightly larger than $\chi$ for $H//c$.
In MLH-\liruh, a magnetic susceptibility shows a broad peak around $T_\mathrm{N} \sim 3.6$ K, which is considered to be an AF transition.
Surprisingly, an AF transition is fully suppressed at least down to 2 K in BLH-\liruh.
From a Curie-Weiss fit for intercalated samples with a function of $\chi = \left ( 1-x \right ) C/(T-\theta_\mathrm{CW})$, the $\theta_\mathrm{CW}$ and $\mu_{eff}$ values are $\theta_\mathrm{CW} = 16$ K and $\mu_{eff} =  1.4~\mu_\mathrm{B}/\mathrm{Ru}^{3+}$ for MLH-\liruh~and $\theta_\mathrm{CW} = -15$ K and $\mu_{eff}=1.6~\mu_\mathrm{B}/\mathrm{Ru}^{3+}$ for BLH-\liruh, respectively, which indicates that the ferromagnetic interaction in \ru~is changed to a weak AF interaction owing to the intercalation.

\section{Discussion}
We now discuss electronic states realized in the Li- and \mizu- intercalated \ru.
The formal valence of Ru in MLH- and BLH- \liruh~$(x\approx0.56)$ is +2.44, so that there are roughly equal number of Ru$^{3+}$ ions with the $(4d)^{5}$ electron configuration $(J_{eff}=1/2)$ and Ru$^{2+}$ ions with the $(4d)^{6}$ electron configuration $(J_{eff}=0)$.
In terms of the band picture, this correspond to the quarter-filled $J_{eff} =1/2$ bands, which is in stark contrast to the half-filled $J_{eff}= 1/2$ bands in \ru. 
The doped electron carriers are expected to conduct smoothly in the system; however, this is not the case. 
The reasons why MLH- and BLH- \liruh~does not show a metallic behavior is likely related to the first-order transition at \ts. 
Taking into account that the number of populated Ru$^{2+}$ and Ru$^{3+}$ ions is almost equal on the bipartite honeycomb lattice, we consider that a charge order with the alternate arrangement of Ru$^{2+}$ and Ru$^{3+}$ ions occurs below \ts.
The rapid increase of $\rho$ below \ts~is consistent with a formation of a charge order. 
If there are relevant fluctuations far above \ts, a non-metallic behavior of intercalated samples at room temperature is also well accounted for. 
We note that the similar scenario is also proposed for K-coated \ru, where photoemission spectra exhibit a gap-like feature at low temperatures \cite{PRM.1.052001}. 
In the charge ordered state, one set of Ru$^{2+}$ and Ru$^{3+}$ ions forms a triangular lattice, and the inversion symmetry is broken.
Comparing to \ts~for MLH- and BLH- \liruh, the former is $\sim 50$ K higher than the latter.
That is, the temperature where a charge order occurs is greatly different between MLH- and BLH- \liruh~in spite of the fact that these two samples have the same electron configuration.
In BLH-system, Li ions are sandwiched between neutral \mizu~layers, which results in the shielding of Coulomb potential of Li ions.
This may relate with the lower \ts~in BLH-\liruh~than that in MLH-\liruh. 

We next discuss the mechanism of the suppressed AF order in the intercalated \ru.
Because Ru$^{2+}$ ions with $(4d)^6$ electron configurations are nonmagnetic, and the doping level exceeds a percolation limit of a honeycomb lattice 0.303, it is quite reasonable to expect the suppression of the AF order. 
More importantly, in the charge-ordered state, magnetic interactions across the nearest-neighbor Ru sites does not work, since one of two adjacent Ru sites is occupied by a nonmagnetic Ru$^{2+}$ ion. 
As a consequence, not only Kitaev-type ferromegneic interaction but also the so-called $\Gamma$ term as a source of magnetic anisotropy does not work effectively, leading to an isotropic spins.
Instead, the next-nearest-neighbor interactions are expected to be dominant in the charge ordered state. 
Therefore, the AF transition at low temperatures in MLH-\liruh~is originating from exchange interactions on a Ru$^{3+}$ triangular lattice.
One plausible candidate of the AF structure in MLH-\liruh~is the $120^\circ$ structure, which hosts the left-handed and right-handed chirality. 
It should be noted that \tn~for BLH-\liruh~is lower than that for MLH-\liruh, while the Li contents are the same in these two samples. 
This indicates that \tn~depends on not only electronic states among honeycomb layer of Ru ions but also interlayer distances. 
As shown in the inset of Fig. \ref{fig:chi}(d), the longer the interlayer distance is, the lower magnetic transition temperature is; this suggests that the interaction between Ru-Cl layers is a origin of the AF transition in MLH-\liruh, and well explains the absence of the magnetic order in BLH-\liruh.
For realizing the electron-doped Kitaev spin liquid, it is important to control the Li content precisely and clarify whether the Kitaev-like correlations remain or not in such a system.

\section{Summary}
In summary, we successfully prepare hydrated and Li-intercalated $\alpha$-\ru, \liruh, ~by using a soft chemical technique.
We found two kinds of crystal structures; one is MLH-\liruh, the other is BLH-\liruh.
The interlayer distance between Ru-Cl layers for MLH- and BLH-\liru~is 1.4-1.9 times larger than that for pristine \ru. 
MLH- and BLH- \liruh~do not show a metallic behavior in the resistivity curves, while a roughly half of Ru sites changes from Ru$^{3+}$ to Ru$^{2+}$.
We consider that this is due to a formation of a charge order at \ts~where a temperature hysteresis in resistivity curves and a rapid increase of the resistivity are observed.
The magnetic susceptibility measurements reveal that MLH-\liruh~shows an antiferromagnetic transition at $T_\mathrm{N}=3.61$ K and that an antiferromagnetic order is suppressed at least down to 2 K in BLH-\liruh, which suggests that the antiferromagnetic transition is sensitive for an electronic state of Ru and an interlayer distance.


\providecommand{\noopsort}[1]{}\providecommand{\singleletter}[1]{#1}

\   \\ 

\noindent Acknowledgements

We would like to thank Fuyuki Sakamoto at Tohoku University for his help in the ICP analysis, and Yukitoshi Motome at the University of Tokyo and Joji Nasu at Yokohama National University for fruitful discussions. 

This work was supported by JSPS KAKENHI Grant Numbers 16K17732, 17H05474, 18H01159, 18H04302, 18K03531, and 19H04685.  \\

\end{document}